\documentclass[12pt]{article}%
\usepackage{putex}
\usepackage{graphicx}
\usepackage{epstopdf}
\usepackage{makeidx}
\usepackage{multicol}
\usepackage{geometry}
\usepackage{amsfonts}
\usepackage{amssymb}
\usepackage{amsmath}%
\begin{document}

\date{November, 2004}

\preprint{hep-th/0411001 \\ PUPT-2141 \\ HUTP-04/A040}%

\institution{PU}%
{Joseph Henry Laboratories, Princeton University, Princeton, NJ 08544}%

\institution{HarvardU}{Jefferson Laboratories, Harvard University, Cambridge,
MA 02138}%

\title{Thermodynamics of R-charged Black Holes in AdS$_{5}%
$ From Effective Strings}%
%

\authors{Steven S. Gubser\worksat{\PU,}\footnote{e-mail: {\tt
ssgubser@princeton.edu}}
and Jonathan J. Heckman\worksat{\HarvardU,}\footnote{e-mail: {\tt
jheckman@fas.harvard.edu}}}%

\abstract{It is well known that the thermodynamics of certain near-extremal
black holes in asymptotically flat space can be lifted to an effective string
description created from the intersection of D-branes. In this paper we
present evidence that the semiclassical thermodynamics of near-extremal
R-charged black holes in $AdS_{5}\times S^{5}$ is described in a similar
manner by effective strings created from the intersection of giant gravitons
on the $S^{5}$. We also present a free fermion description of the
supersymmetric limit of the one-charge black hole, and we give a crude catalog
of the microstates of the two and three-charge black holes in terms of
operators in the dual conformal field theory.}%

\maketitle

\section{Introduction}

\label{INTRODUCTION}

Little exact information is available for black holes in $AdS_{p}$ with $p>3$.
For example, there is a $3/4$ mismatch between the free energy of an uncharged
black hole in $AdS_{5}$ and the free-field calculation of free energy in the
dual gauge theory \cite{gkPeet}. This mismatch is thought to be due to the
strong interactions that are present in the gauge theory in the regime where
supergravity calculations are reliable. Calculations both in supergravity
\cite{gkTseytlin} and field theory \cite{ft} support the view that
$f(g_{YM}^{2}N_{c})\equiv-F/N_{c}^{2}VT^{4}$ smoothly interpolates between its
free field value at $g_{YM}^{2} N = 0$ and $3/4$ this value at $g_{YM}^{2} N =
\infty$.

Novel results on the AdS/CFT correspondence \cite{juanAdS,gkPol,witHolOne}
have been obtained by studying states with some large quantum number, like
R-charge \cite{bmn} or angular momentum \cite{gkPolTwo}. Many of these results
have yielded new exact comparisons between string theory and a dual gauge
theory. In light of all this progress, it is tempting to reconsider black
holes in $AdS_{5}$ with one, two, or three large charges. On the supergravity
side, these black holes are solutions of $d=5$, $\mathcal{N}=8$ gauged
supergravity \cite{bcsNONextreme}. The three charges correspond to the
$U(1)^{3}$ Cartan subalgebra of the gauge group $SO(6)$. In $AdS_{5}\times
S^{5}$, these backgrounds can be described as black holes with three
independent angular momenta in the $S^{5}$ directions. In asymptotically flat
space, a limit of these solutions describes D3-branes with angular momenta in
the three independent planes orthogonal to the brane world-volumes. In the
dual gauge theory, these charged black holes correspond to a thermal bath,
uniform on the $S^{3}$ spatial slice of the boundary of $AdS_{5}$, and
carrying three R-charges: the R-symmetry group, of course, is the same $SO(6)$
that describes rotations on $S^{5}$.

These R-charged black holes have been studied extensively: see for example
\cite{bcsNONextreme,bcsNaked,cgOne,cgTwo,MyersCatastrophic,SuperStars,bpz}.
They have some similarities to the three-charge black holes in asymptotically
flat space which figured so prominently in the string theory counting of black
hole microstates \cite{sv,cm,hs}. The purpose of this paper is to examine the
extent to which similar methods, based on intersecting D-branes, can be used
to account for the semiclassical thermodynamics of the R-charged black holes
in $AdS_{5}$. Our main result is that, under certain plausible assumptions, in
the regime where at least one charge is small compared to the others,
intersecting giant gravitons do indeed give a microstate counting that agrees
with the semiclassical results. It is interesting to note that for the
three-charge case, even at zero temperature, a horizon breaks supersymmetry:
this is in contrast to the asymptotically flat space case, where BPS
configurations exist with a macroscopic horizon.

The organization of the rest of the paper is as follows. In section~\ref{BH}
we describe the R-charged black holes as the classical backgrounds in $d=5$,
$\mathcal{N}=8$ supergravity. After establishing the relative sizes of the CFT
charges to $N_{c}$ in section~\ref{BOUNDS}, in section~\ref{EFFECTIVE} we show
that the semiclassical thermodynamics of the black holes is reproduced (with
the aforementioned caveats) by effective strings created by the intersection
of distributions of giant gravitons. In section~\ref{FREEFERMIONS} we exhibit
a free fermion description of the supersymmetric limit of the one-charge black
hole. We conclude in section~\ref{CONCLUDE} with some further observations
about these black holes and their field theory duals.

The results in this paper are based in part on the senior thesis of JJH
\cite{hthesis}.

\section{R-charged black holes in $AdS_{5}$}

\label{BH}

In $d=5$, $\mathcal{N}=8$ gauged type IIB supergravity, the metric for an
R-charged black hole in $AdS_{5}$ is given by
\begin{equation}
ds^{2}=g_{\mu\nu}dx^{\mu}dx^{\nu}=-H^{-2/3}fdt^{2}+H^{1/3}\left(  f^{-1}%
dr^{2}+r^{2}d\Omega_{3,k}^{2}\right)  \label{metric}%
\end{equation}
where $d\Omega_{3,k}^{2}$ is the metric on an $S^{3}$, and
\begin{align}
H_{i} &  =1+\frac{q_{i}}{r^{2}}\label{HiDef}\\
H &  =\text{ }\underset{i=1}{\overset{3}{\prod}}\left(  1+\frac{q_{i}}{r^{2}%
}\right)  \label{HDef}\\
f &  =k-\frac{\mu}{r^{2}}+\frac{r^{2}}{L^{2}}H\,.\label{fDef}%
\end{align}
Here $L$ is the radius of $AdS_{5}$, and $k$ controls the size of the $S^{3}$.
\ The value $k=1$ corresponds to an $S^{3}$ of radius $1$, and the value $k=0$
corresponds to the conformal completion of $%
\mathbb{R}
^{3}$. In all that follows, we assume $k=1$. The charges $q_{i}$ are related
to the three independent angular momenta in $AdS_{5}\times S^{5}$. The
parameter $\mu\geq0$ describes departures from extremality: at $\mu=0$
supersymmetry is restored. The mass of the black hole is given by the formula:%
\begin{equation}
M=\frac{\pi}{4G_{5}}\left(  \frac{3}{2}\mu+q_{1}+q_{2}+q_{3}\right)
\,\label{massformula}%
\end{equation}
where $G_{5}$ is the five dimensional Newton constant. The $q_{i}$ are not
conserved charges, but they are simply related to conserved charges $\tilde
{q}_{i}$: \eqn{physcharge}{
\tilde{q}_{i}=\sqrt{q_{i}(\mu+q_{i})} \,.
} The scalar potentials for the three $U(1)$ gauge fields are
\begin{equation}
A_{t}^{i}=\frac{\tilde{q}_{i}}{r^{2}+q_{i}}\,,
\end{equation}
and there are also three real scalar fields (parametrizing directions in the
$E_{6,6}/USp(8)$ coset that describes the 42 scalar fields in $d=5$,
$\mathcal{N}=8$ gauged supergravity) \cite{bcsNONextreme,cgOne}
\begin{equation}
X^{i}=H_{i}^{-1}\left(  \prod_{i=1}^{3}H_{i}\right)  ^{1/3}\,.
\end{equation}

In general, we will work close enough to extremality ($\mu=0$) that the
difference between $q_{i}$ and $\tilde{q}_{i}$ is negligible. Note however
that for at least two $q_{i}$ non-zero, one cannot take $\mu$ arbitrarily
small without losing the horizon altogether: the radial location of the outer
horizon $r_{+}$ is the largest double zero of $g_{tt}$, which leads to
\begin{equation}
f\left(  r_{+}\right)  =1-\frac{\mu}{r_{+}^{2}}+\frac{r_{+}^{2}}{L^{2}}\left(
1+\frac{q_{1}}{r_{+}^{2}}\right)  \left(  1+\frac{q_{2}}{r_{+}^{2}}\right)
\left(  1+\frac{q_{3}}{r_{+}^{2}}\right)  =0. \label{fhor}%
\end{equation}
Let us refer to the smallest value of $\mu$ that leads to a horizon as
$\mu_{\mathrm{crit}}$. For all $\mu> \mu_{\mathrm{crit}}$, there is a regular
horizon with finite temperature. For all $\mu< \mu_{\mathrm{crit}}$, there is
a naked singularity at $r=0$. For $\mu=0$ (the supersymmetric case) the
configurations (\ref{metric}) have been referred to as ``superstars.''

For $\mu=\mu_{\mathrm{crit}}$, $f(r)$ has a double zero at $r=r_{+}$: thus
$f(r_{+}) = f^{\prime}(r_{+}) = 0$. These two conditions can be expressed more
simply as \eqn{ExpressDouble}{\seqalign{\span\TC}{
L^2 + q_1 + q_2 + q_3 - {q_1 q_2 q_3 \over r_+^4} + 2 r_+^2 = 0  \cr
\mu_{\rm crit} L^2 = -r_+^4 + q_1 q_2 + q_1 q_3 + q_2 q_3 + {2 q_1 q_2 q_3 \over r_+^2} \,.
}} The first of these may be solved explicitly to give $r_{+}/L$ in terms of
$q_{i}/L^{2}$, and then the second may be used to express $\mu/L^{2}$ in terms
of $q_{i}/L^{2}$. The resulting expressions are not simple, so we will not
record them here. Let us however quote some special limits.

\begin{itemize}
\item $q_{2}=q_{3}=0$. In this case, when $\mu=0$, there is a naked
singularity in the five-dimensional solution at $r=0$, and for $\mu>0$, there
is a horizon at \eqn{SingleChargeRp}{
r_{+(1)}^2 = {\mu L^2 \over L^2+q_1} + O(\mu^2) \,.
} When the black hole is arbitrarily close to criticality, the outer horizon
has vanishingly small size.

\item $q_{3}=0$. When $\mu<\mu_{\mathrm{crit}}=q_{1}q_{2}/L^{2}$ there is a
naked singularity at $r=0$. For $\mu\geq\mu_{\mathrm{crit}}$, there is a
horizon at \eqn{TwoChargeRp}{
r_{+(2)}^2 &= {L^2+q_1+q_2 \over 2} \left( -1 +
\sqrt{1 + 4L^2 {\mu-\mu_{crit} \over (L^2+q_1+q_2)^2}} \right)  \cr
&= L^4 {\mu-\mu_{\rm crit} \over L^2+q_1+q_2} + O[(\mu-\mu_{\rm crit})^2] \,,
} which is vanishingly small as the black hole approaches criticality.

\item $0<q_{i}\ll L^{2}$. There is a naked singularity at $r=0$ when
\eqn{muCritThree}{
\mu < \mu_c = 2 {\sqrt{q_1 q_2 q_3} \over L} +
{q_1 q_2 + q_1 q_3 + q_2 q_3 \over L^2} +
O(q_i^{5/2}) \,.
} For $\mu>\mu_{\mathrm{crit}}$, there is a regular horizon at
\eqn{ThreeChargeRp}{
r_{+(3)}^2 &= {\sqrt{q_1 q_2 q_3} \over L} +
{(q_1 q_2 q_3)^{3/4} \sqrt{\mu-\mu_c} \over \sqrt{L}} \,.
} In contrast to the previous cases, as the black hole approaches criticality,
it maintains a finite sized outer horizon.
\end{itemize}

The semiclassical entropy and temperature of the black holes under
consideration is \eqn{EntTemp}{
S &= \frac{A}{4G_{5}} = \frac{2\pi^{2}}{4G_{5}}\sqrt{\prod_{i=1}^3
(r_{+}^{2}+q_{i})}  \cr
\beta &= {1 \over T} = L^{2}\frac{2\pi r_{+}^{2}
\sqrt{\prod_{i=1}^3 (r_{+}^{2}+q_{i})}}{2r_{+}^{6}+r_{+}^{4} \left( L^{2}+\sum q_{i}\right) - \prod_{i=3}^3 q_{i}} \,.
} We now apply the above formulas to the cases discussed above:

\begin{itemize}
\item $q_{2}=q_{3}=0$. In this case, \eqn{STOne}{
S_{(1)} &= {2\pi^2 \over 4 G_5} r^2_{+(1)} \sqrt{q_1} \approx
\frac{2\pi^{2}}{4G_{5}}\mu {\sqrt{q_{1}} \over 1 + q_1/L^2}  \cr
\beta_{(1)} &\approx 2\pi {\sqrt{q_{1}} \over 1 + q_1/L^2} \,,
} where the approximate equalities are accurate to leading order in small
$\mu$.

\item $q_{3}=0$. Assuming also that $q_{2}$ and $q_{3}$ are much less than
$L^{2}$, one obtains \eqn{STTwo}{
S_{(2)} &\approx \frac{2\pi^{2}}{4G_{5}}r_{+(2)}\sqrt{q_{1}q_{2}}
\approx \frac{2\pi^{2}}{4G_{5}}\sqrt{\left( \mu-\mu_{\text{crit}}\right)}
\sqrt{q_{1}q_{2}}  \cr
\beta_{(2)} &\approx\frac{2\pi\sqrt{q_{1}q_{2}}}{r_{+(2)}}
\approx \frac{2\pi\sqrt{q_{1}q_{2}}}
{\sqrt{\left( \mu-\mu_{\text{crit}}\right)}} \,,
} where in each line the second approximate equality comes from expanding to
non-trivial leading order in small $\mu-\mu_{\mathrm{crit}}$.

\item All charges non-zero, and much less than $L^{2}$. In this case, the
entropy is given by: \eqn{STThree}{
S_{(3)} &\approx \frac{2\pi^{2}}{4G_{5}}\sqrt{q_{1}q_{2}q_{3}}
} At criticality, the temperature vanishes. As $\mu$ approaches $\mu
_{\mathrm{crit}}$, $\beta$ becomes: \eqn{STThree}{
\beta_{(3)} &\approx \frac{\pi}{L}\sqrt{\frac{q_{1}q_{2}q_{3}}{\mu-\mu_{crit}}} \,.
}
\end{itemize}

\section{CFT\ dictionary entries}

\label{CFT}

We now recast the parameters of the above black holes in terms of dual CFT
dictionary entries. After also establishing the dictionary entries for the
parameters $\mu$ and $\mu_{\text{crit}}$ we state dual CFT formulas for the
entropy and temperature. \ Some of the results of this section overlap with
the discussion in \cite{SilvaGGs}.

Although small in comparison to $AdS_{5}$, black holes which can be probed by
supergravity are still macroscopic objects and as such the CFT operators which
describe the black hole will have large scaling dimension $\Delta=ML$. \ Using
the black hole mass formula of equation \ref{massformula} thus yields:%
\begin{equation}
\Delta=\frac{\pi L}{4G_{5}}\left(  \frac{3}{2}\mu+q_{1}+q_{2}+q_{3}\right)  .
\label{mass}%
\end{equation}
In the limit $\mu=q_{2}=q_{3}=0$ the CFT dual of the singularity becomes a
BPS\ object with scaling dimension:%
\begin{equation}
\Delta_{\text{BPS(1)}}=\frac{\pi L}{4G_{5}}q_{1}. \label{GWAH}%
\end{equation}
BPS\ operators in the dual CFT with corresponding $U(1)\subset SU(4)_{R}$
charge $J_{1}$ must satisfy the single charge BPS bound:%
\begin{equation}
\Delta_{\text{BPS(1)}}-J_{1}=0.
\end{equation}
Equation \ref{GWAH} then implies:%
\begin{equation}
J_{1}=\frac{\pi L}{4G_{5}}q_{1}.
\end{equation}
Using the AdS/CFT\ dictionary entry $2G_{5}N_{c}^{2}=\pi L^{3}$ we find
\begin{equation}
\frac{J_{1}}{N_{c}^{2}}=\frac{q_{1}}{2L^{2}}.
\end{equation}
Extrapolating to the other R-charges thus yields
\begin{equation}
\frac{J_{i}}{N_{c}^{2}}=\frac{q_{i}}{2L^{2}} \label{relation}%
\end{equation}
for $i=1,2,3$. These relations fully determine the corresponding dictionary
entry for $\mu$. Equations \ref{mass} and \ref{relation} imply
\begin{equation}
\Delta-\left(  J_{1}+J_{2}+J_{3}\right)  =\frac{\pi L}{4G_{5}}\frac{3}{2}\mu
\end{equation}
or,%
\begin{equation}
\frac{4}{3}\frac{\left(  \Delta-\left(  J_{1}+J_{2}+J_{3}\right)  \right)
}{N_{c}^{2}}=\frac{\mu}{L^{2}}. \label{murelation}%
\end{equation}
The critical operator dimension $\Delta_{\text{crit}}$ necessary to form a
horizon is thus related to $\mu_{\text{crit}}$ by:
\begin{equation}
\frac{4}{3}\frac{\left(  \Delta_{\text{crit}}-\left(  J_{1}+J_{2}%
+J_{3}\right)  \right)  }{N_{c}^{2}}=\frac{\mu_{\mathrm{crit}}}{L^{2}}
\label{murelationcrit}%
\end{equation}
so that
\begin{equation}
\frac{4}{3}\frac{\left(  \Delta-\Delta_{\text{crit}}\right)  }{N_{c}^{2}%
}=\frac{\mu-\mu_{\mathrm{crit}}}{L^{2}}%
\end{equation}
where
\begin{equation}
\frac{\mu_{\mathrm{crit}}}{L^{2}}=2\left(  \frac{\left(  2J_{1}\right)
\left(  2J_{2}\right)  \left(  2J_{3}\right)  }{N_{c}^{6}}\right)
^{1/2}+\frac{\left(  2J_{1}\right)  \left(  2J_{2}\right)  +\left(
2J_{1}\right)  \left(  2J_{3}\right)  +\left(  2J_{2}\right)  \left(
2J_{3}\right)  }{N_{c}^{4}}+O\left(  \frac{J_{i}^{5/2}}{N_{c}^{5}}\right)  .
\end{equation}
Solving equation \ref{murelationcrit} for $\Delta_{\text{crit}}$ yields
\begin{equation}
\Delta_{\text{crit}}=\Upsilon+\Delta_{\text{BPS}} \label{beeboba}%
\end{equation}
where by abuse of notation, we have defined $\Delta_{\text{BPS}}\equiv
J_{1}+J_{2}+J_{3}$ and:%
\begin{equation}
\Upsilon=\frac{3}{4}\left(  2\frac{\sqrt{\left(  2J_{1}\right)  \left(
2J_{2}\right)  \left(  2J_{3}\right)  }}{N_{c}}+\frac{\left(  2J_{1}\right)
\left(  2J_{2}\right)  +\left(  2J_{1}\right)  \left(  2J_{3}\right)  +\left(
2J_{2}\right)  \left(  2J_{3}\right)  }{N_{c}^{2}}+O\left(  \frac{J_{i}^{5/2}%
}{N_{c}^{3}}\right)  \right)  .
\end{equation}

With the dictionary entries for the $q_{i}$'s and $\mu$ established, we now
recast all thermodynamic predictions for the temperature and entropy from
supergravity in terms of purely CFT\ quantities. Making the trivial
substitutions, we have for the single, double and triple charge cases:%
\begin{align}
S_{(1)}  &  =\frac{\pi}{N_{c}}\frac{4}{3}\left(  \Delta-\Delta_{\text{crit}%
}\right)  {\frac{\sqrt{2J_{1}}}{1+2J_{1}/N_{c}^{2}}}\label{entropysingle}\\
\beta_{(1)}  &  =\frac{2\pi L}{N_{c}}{\frac{\sqrt{2J_{1}}}{1+2J_{1}/N_{c}^{2}%
}}\label{tsingle}\\
S_{(2)}  &  =\frac{\pi}{N_{c}}\sqrt{\frac{4}{3}\left(  \Delta-\Delta
_{\text{crit}}\right)  }\sqrt{\left(  2J_{1}\right)  \left(  2J_{2}\right)
}\label{sdouble}\\
\beta_{(2)}  &  =\frac{2\pi L}{N_{c}}\frac{\sqrt{(2J_{1})(2J_{2})}}%
{\sqrt{\frac{4}{3}\left(  \Delta-\Delta_{\text{crit}}\right)  }}%
\label{tdouble}\\
S_{(3)}  &  =\frac{\pi}{N_{c}}\sqrt{\left(  2J_{1}\right)  \left(
2J_{2}\right)  \left(  2J_{3}\right)  }\label{striple}\\
\beta_{(3)}  &  =\frac{\pi L}{N_{c}^{2}}\sqrt{\frac{\left(  2J_{1}\right)
\left(  2J_{2}\right)  \left(  2J_{3}\right)  }{\frac{4}{3}\left(
\Delta-\Delta_{\text{crit}}\right)  }}\,. \label{ttriple}%
\end{align}
Except in the single-charge case, we have assumed that all the $J_{i}$ are
much less than $N_{c}^{2}$; also, we have in all cases expanded to the leading
non-trivial order in $\Delta-\Delta_{\mathrm{crit}}$. Although we have
expressed the entropy of the three possible charge configurations solely in
terms of CFT quantities, the CFT\ temperature predictions still depend on $L$,
which appears to be a purely $AdS$ quantity. \ Viewing the four dimensional
gauge theory as a field theory on the boundary of $AdS$, the factor of $L$
sets the energy scale for the $S^{1}$ of the full $S^{3}\times S^{1}$ geometry
of the gauge theory.

It is very tempting to write a general formula which interpolates between the
various charge configurations near extremality, at least when all $J_{i}\ll
N_{c}^{2}$. By inspection of equations (\ref{entropysingle})-(\ref{ttriple}),
the formula which accomplishes this is:%
\begin{align}
S  &  =\frac{\pi}{N_{c}}\sqrt{\underset{i}{\prod}\left(  2J_{i}+\frac{4}%
{3}\left(  \Delta-\Delta_{\text{crit}}\right)  \right)  }\\
\beta &  =\frac{2\pi L}{N_{c}}\sqrt{\underset{i}{\prod}\left(  2J_{i}+\frac
{4}{3}\left(  \Delta-\Delta_{\text{crit}}\right)  \right)  }\frac{\left(
1+\frac{J_{1}J_{2}J_{3}}{2N_{c}}\sqrt{\frac{4}{3}\left(  \Delta-\Delta
_{\text{crit}}\right)  }\right)  }{\left(  1+J_{1}J_{2}J_{3}\right)  \frac
{4}{3}\left(  \Delta-\Delta_{\text{crit}}\right)  }.
\end{align}
As we shall see in the next section, so long as it does not vanish, the
$J_{1}J_{2}J_{3}/N_{c}$ term in the equation for $\beta$ is of order at least
$N_{c}$, justifying the use of the above interpolation formula.

\section{Bounds on the semiclassical regime}

\label{BOUNDS}

We now determine the regime of validity for the semiclassical formulas found
above. The requirement that the near horizon curvature remain small will also
fix the relative sizes of the CFT charges $J_{i}$ and the large number $N_{c}%
$. One particularly useful measure of the near horizon curvature is:%
\begin{equation}
G_{5}^{4/3}C\equiv\left\vert G_{5}^{4/3}R_{\mu\nu}R^{\mu\nu}\right\vert
_{r=r_{+}}\ll1.
\end{equation}
The leading order behavior of $R_{\mu\nu}R^{\mu\nu}$ in the various charge
cases is given by:%
\begin{align}
\text{ }C_{1}  &  =\frac{1}{L^{4}}\frac{16}{3}\frac{q_{1}/L^{2}}{\left(
\mu/L^{2}\right)  ^{2}}+...\\
C_{2}  &  =\frac{1}{L^{4}}\left(  \frac{16}{3}\frac{\left(  q_{1}%
+q_{2}\right)  /L^{2}}{\left(  \left(  \mu-\mu_{\mathrm{crit}}\right)
/L^{2}\right)  ^{2}}+\frac{88}{9}\frac{\left(  q_{1}/L^{2}\right)  \left(
q_{2}/L^{2}\right)  }{\left(  \left(  \mu-\mu_{\mathrm{crit}}\right)
/L^{2}\right)  ^{4}}\right)  +...\\
C_{3}  &  =\frac{44}{L^{4}}\left(  \frac{q_{1}}{L^{2}}\frac{q_{2}}{L^{2}}%
\frac{q_{3}}{L^{2}}\right)  ^{-2/3}+...
\end{align}
Using the dictionary entries established in the previous section, the
requirement that the near horizon curvature remain small yields the following
constraints on the parameters of the semiclassical regime in the single,
double and triple charge cases respectively:%
\begin{align}
N_{c}^{2/3}  &  \gg\left(  \frac{J_{1}}{\left(  \Delta-\Delta_{\text{crit}%
}\right)  ^{2}}\right) \label{relsingle}\\
N_{c}^{2/3}  &  \gg\left(  \frac{\left(  J_{1}+J_{2}\right)  }{\left(
\Delta-\Delta_{\text{crit}}\right)  ^{2}}+\frac{J_{1}J_{2}N_{c}^{2}}{\left(
\Delta-\Delta_{\text{crit}}\right)  ^{4}}\right) \label{reldouble}\\
N_{c}^{2}  &  \ll J_{1}J_{2}J_{3}. \label{reltriple}%
\end{align}
Where in the above bounds we have dropped unimportant factors of order unity.
\ Whereas the single and double charges cases are bounded above by an
appropriate power of $N_{c}$, in the triple charge case we instead find a
lower bound.

\section{Effective strings from giant gravitons}

\label{EFFECTIVE}

We now present evidence that an effective string created by a distribution of
giant gravitons on the spherical factor of $AdS_{5}\times S^{5}$ has the same
thermodynamics as a near-extremal R-charged black hole with charges satisfying
$q_{i}/L^{2} \ll1$. (By near-extremal we mean having energies only slightly
above the threshold to produce a regular horizon). After analyzing the single
charge case in section~\ref{ONECHARGE}, we consider in
sections~\ref{TWOCHARGE} and~\ref{THREECHARGE} more general giant graviton
distributions on the $S^{5}$ and show that the thermodynamics of the effective
strings produced in these cases also match well to the two and three charge
black holes.

When the extremality parameter $\mu=0$, the black hole becomes an uncloaked
singularity or \textquotedblleft superstar\textquotedblright%
\ \cite{bcsNaked,SuperStars}. Recent work has shown that single charge
superstars in $AdS_{5}$ are well-described by a distribution of gravitons
smeared out along one of the equators of the $S^{5}$ \cite{SuperStars}.
Following \cite{SuperStars}, we parametrize the $S^{5}$ of radius $L$ by
angular coordinates $\theta_{1},\theta_{2},\phi_{1},\phi_{2}$ and $\phi_{3}$
where the three distinct equatorial directions are specified by the three
$\phi$ angles. Assuming that the giant gravitons orbit the $S^{5}$ along the
$\phi_{1}$ equator, their distribution $dn_{1}/d\theta_{1}$ in the $\theta
_{1}$ direction is determined by integrating over the angular dependence of
the five-form flux in the other directions of the $S^{5}$. This distribution
is computed in \cite{SuperStars} with the result:%
\begin{equation}
\frac{dn_{1}}{d\theta_{1}}=N_{c}\frac{q_{1}}{L^{2}}\sin2\theta_{1}.
\end{equation}
The total number of such giant gravitons is obtained by integrating over all
values of $\theta_{1}$ for which the five-form flux is positive:%
\begin{equation}
n_{1}=\int_{0}^{\pi/2}\frac{dn_{1}}{d\theta_{1}}d\theta_{1}=N_{c}\frac{q_{1}%
}{L^{2}}.
\end{equation}
As determined in \cite{Invasion}, the angular momentum of a single giant
graviton orbiting in the $\phi_{1}$ direction and at azimuthal angle
$\theta_{1}$ is:%
\begin{equation}
p_{\phi_{1}}=N_{c}\sin^{2}\theta_{1}%
\end{equation}
which implies that the total angular momentum of the giant graviton
distribution is \cite{SuperStars}:%
\begin{equation}
P_{\phi_{1}}=\int_{0}^{\pi/2}\frac{dn_{1}}{d\theta_{1}}N_{c}\sin^{2}\theta
_{1}d\theta_{1}=N_{c}^{2}\frac{q_{1}}{2L^{2}}.
\end{equation}
Using the dictionary entries previously established, we thus find that the
number of giant gravitons is exactly:%
\begin{equation}
n_{1}=\frac{2J_{1}}{N_{c}}%
\end{equation}
with net angular momentum:%
\begin{equation}
P_{\phi_{1}}=J_{1}.
\end{equation}
Based on the \textquotedblleft stringy exclusion principle\textquotedblright%
\ found in \cite{ExclusionPrinciple}, it has been shown that a single giant
graviton can have maximal angular momentum $N_{c}$ before it inflates to the
full size of the $S^{5}$ \cite{Invasion}. Hence, in addition to the small
black hole constraint $J_{1}\ll N_{c}^{2},$ we must also have $J_{1}\gtrsim
N_{c}$ in order to have a finite number of giant gravitons.

\subsection{One distribution of giant gravitons}

\label{ONECHARGE}

In this section we present evidence that the transverse intersection of a
single D3-brane with the superstar distribution of giant gravitons described
above creates an effective string which explains the thermodynamic properties
of the single charge black hole. To remain in the regime well-described by
both supergravity and the giant graviton distribution, the single charge
supergravity bound of equation~(\ref{relsingle}),%
\begin{equation}
\left(  \frac{J_{1}}{\left(  \Delta-J_{1}\right)  ^{2}}\right)  \ll
N_{c}^{2/3}\,
\end{equation}
implies:%
\begin{equation}
J_{1}+\frac{\sqrt{J_{1}}}{N_{c}^{1/3}}\ll\Delta\label{singlechargedimdim}%
\end{equation}
where we have used the fact that in the single charge case $\Delta
_{\text{crit}}=J_{1}$. Since $J_{1}$ scales as $N_{c}^{1+\delta}$ for some
$0<$ $\delta<1$, $J_{1}$ dominates the lower bound of the above inequality.
\ Thus, although $\Delta$ will necessarily be above the BPS bound, it can
still scale as $J_{1}N_{c}^{\varepsilon}$ for small $\varepsilon>0$.

The value of the central charge of the effective string created from the
intersection of $2J_{1}/N_{c}$ giant gravitons and a single transverse
D3-brane follows from the appendix:%

\begin{equation}
c_{\text{eff(1)}}=12\frac{J_{1}}{N_{c}}.
\end{equation}
Matching the Hagedorn temperature of the effective string to the Hawking
temperature of the black hole yields:%
\begin{equation}
\frac{L^{2}}{N_{c}}=\alpha_{\text{eff}}^{\prime}%
\end{equation}
so that the tension of the effective string becomes:%
\begin{equation}
\tau_{\text{eff}}=\frac{1}{2\pi\alpha_{\text{eff}}^{\prime}}=\frac{N_{c}}{2\pi
L^{2}}.
\end{equation}
Treating the effective string as a real curve inside the giant graviton
$S^{3}$, the corresponding volume of a real codimension two $S^{2}$ is given
by:%
\begin{equation}
\tau_{\text{D3}}Vol(S^{2})=\tau_{\text{eff}}%
\end{equation}
so that:%
\begin{equation}
Vol(S^{2})=\pi L^{2} \label{StwoOneCharge}%
\end{equation}
which is exactly a quarter the size of the maximum possible $S^{2}$ of volume
$4\pi L^{2}$. Hence, the radius of the $S^{2}$ never exceeds the maximal value
of $L$. \ This gives provisional evidence that the intersecting giant graviton
picture considered here gives a proper description of single charge black
holes in $AdS_{5}$: such a black hole slightly above extremality corresponds
to an ensemble of highly excited effective strings at their Hagedorn temperature.

\subsection{Two distributions of giant gravitons}

\label{TWOCHARGE}

As we observed around \TwoChargeRp, the two-charge black hole in $AdS_{5}$
acquires a regular horizon finitely far from the supersymmetric limit: the
non-extremality parameter $\mu$ must be greater than or equal to
$\mu_{\text{crit}}\equiv q_{1}q_{2}/L^{2}$. (Recall that we require $q_{i}\ll
L^{2}$). This makes it less clear that string theory should provide a correct
accounting of the microstates of a black hole with $\mu$ slightly larger than
$\mu_{\text{crit}}$. Nevertheless, we will argue here that an effective string
picture once again succeeds in reproducing the thermodynamics up to an overall
factor of order unity which we have been unable to determine.

In contrast to the one-charge case where the effective strings are highly
excited and the temperature of the black hole close to extremality is
identified with the Hagedorn temperature of the strings, here the picture is
that an effective string created by the intersection of two distributions of
giant gravitons is stretched over a length $L_{\text{eff}}$ and has
excitations running along it which can be described in terms of a conformal
field theory. If this conformal field theory has central charge $c_{\text{eff}%
}$, then up to factors of order unity the free energy of the excitations we
have described is \eqn{EnergyDescribed}{
F = -c_{\text{eff}} L_{\text{eff}} T^2 \,,
} where $T$ is the temperature. The entropy, which we assume is carried
entirely by the excitations of the conformal field theory on the effective
string, satisfies \eqn{EntropySatisfies}{
{S \over T} = -{{\partial F/\partial T} \over T} =
2c_{\text{eff}} L_{\text{eff}} \,,
} where we have again neglected factors of order unity. Let us use the
supergravity results \eno{sdouble} and \eno{tdouble} for $S$ and $T$ in the
two-charge case to estimate \eqn{cLestimate}{
2c_{\text{eff}} L_{\text{eff}} =
8\pi^{2}L\frac{J_{1}}{N_{c}}\frac{J_{2}}{N_{c}} \,.
}

The effective string arises from the intersection of two distributions of
giant gravitons orbiting different equators of the $S^{5}$, with $2J_{1}%
/N_{c}$ giant gravitons along one equator and $2J_{2}/N_{c}$ along the other.
As determined in the appendix, the central charge of the effective string is
\eqn{ceffTwoCharge}{
c_{\text{eff}(2)}=12\frac{J_{1}}{N_{c}}\frac{J_{2}}{N_{c}}.
}Combining \cLestimate\ and \ceffTwoCharge, the effective string has length
\eqn{LeffTwoCharge}{
L_{\text{eff}(2)}=\frac{\pi}{6}(2\pi L) \,,
} which is a well behaved length for a circle on the $S^{5}$ of radius $L$. As
with the final result \eno{StwoOneCharge} of section~\ref{ONECHARGE}, the main
point is that elementary reasoning based on simple string theory pictures give
the correct number of microstates to understand the semiclassical black hole entropy.

\subsection{The three-charge case in terms of the effective string}

\label{THREECHARGE}

The successful description of black hole microstates of the double charge
black holes suggests that the three-charge case should not be much different,
at least in a limit where the third charge is much smaller than the first two.
The reason is that (again in analogy with the asymptotically flat space cases)
the third charge can then be regarded as momentum along the string in one direction.

As with the two-charge case, one must proceed finitely above the
supersymmetric mass limit to achieve a regular horizon. We will be unable to
give any account of this finite mass gap, and we will refer only to entropy
and conserved charges. We note however that for $q_{3}\ll q_{1}q_{2}/L^{2}$,
the mass gap becomes the same for the two and three-charge cases (compare
\eno{muCritThree} with the discussion above \eno{TwoChargeRp}). This would
thus be a good starting point for understanding why there is some mass gap in
the two charge case.

We propose to describe the microstates behind the zero-temperature horizon
that forms exactly at $\mu=\mu_{\text{crit}(3)}$ in terms of states of the
effective string where there are excitations moving only in one direction.
Such states can indeed have finite entropy and zero-temperature. Their entropy
comes from Cardy counting: \eqn{Cardy}{
S_{\text{Cardy}}=2\pi\sqrt{\frac{nc_{\text{eff}}}{6}} \,,
} where $n$ is the level of the excitations moving in one direction: thus
$n=J_{3}$. According to the discussion above and the computation performed in
the appendix, we should use the same $c_{\text{eff}}$ as in the two-charge
case:
\begin{equation}
c_{\text{eff}(3)}=c_{\text{eff}(2)}=12\frac{J_{1}}{N_{c}}\frac{J_{2}}{N_{c}%
}\,.
\end{equation}
Comparing \eno{Cardy} to the result \eno{striple}, one sees that the level of
the effective string would need to be:%
\begin{equation}
n=J_{3}\,.
\end{equation}
This agrees with our previous claim that $n$ should be identified with $J_{3}$.

Just as the individual graviton could have maximal angular momentum $N_{c}$,
for an effective string picture to be accurate, the highest mode excitation
$n$ must be bounded above by $N_{c}$. To describe the case where all three
$q_{i}$ are comparable, one would need to consider the quantum mechanics of
three distributions of giant gravitons. This appears to be a formidable problem.

In summary, the account of microstates based on intersecting giant gravitons
and effective strings agrees quite well with the entropy of the corresponding
charged black holes in $AdS_{5}$. The main puzzle remaining is the mass gap in
the two and three charge cases between the supersymmetric bound on the mass
and the minimum mass needed to form a regular horizon. Qualitatively, what has
to happen in the CFT is that there are very few operators with $\Delta
-J_{1}-J_{2}-J_{3}$ less than this mass gap, but for larger ``twist'' there
are such a profusion of operators that they correspond to macroscopic horizons
in $AdS_{5}$.

\section{The one-charge black hole and free fermions}

\label{FREEFERMIONS}

\label{Discussion}In light of recent results \cite{Berenstein,LLM} on the
description of supersymmetric states asymptotic to $AdS_{5}\times S^{5}$ with
one non-zero angular momentum, it is interesting to inquire how the free
fermion picture might elucidate the physics of the single charge case. For
$\mu=0$ and only $q_{1}\neq0$, the solution (\ref{metric}) with $k=1$, when
lifted to ten dimensions, can be recast in the form described by Eq.~(2.5) of
\cite{LLM}.\footnote{The explicit form of the solution that we present here is
known to O.~Lunin and J.~Maldacena \cite{MLunpublished}.}

The metric in ten dimensions \cite{SuperStars} is \eqn{MTsoln}{
ds^2 &= \sqrt{\Delta} \left[ -{f \over H} dt^2 +
{dr^2 \over f} + r^2 d\Omega_3^2 \right]  \cr
&\quad{} +
{1 \over \sqrt{\Delta}}
H \left[ L^2 d\mu_1^2 + \mu_1^2 \left( L d\phi_1 +
\left( {1 \over H} - 1 \right) dt \right)^2 \right]  \cr
&\quad{} +
{1 \over \sqrt{\Delta}} \sum_{i=2}^3
L^2 (d\mu_i^2 + \mu_i^2 d\phi_i^2) \,,
} where the $\mu_{i}$ are constrained by $\sum_{i=1}^{3}\mu_{i}^{2}=1$, the
functions $f$ and $H$ are as given in (\ref{HDef}) and (\ref{fDef}), and
$\Delta=H-\mu_{1}^{2}(H-1)$. The second 3-sphere, $\tilde{S}^{3}$ in the
notation of \cite{LLM}, is in the hyperplane parametrized as $(\mu_{2}\cos
\phi_{2},\mu_{2}\sin\phi_{2},\mu_{3}\cos\phi_{3},\mu_{3}\sin\phi_{3})$, and
its radius (from the metric \MTsoln) is $L\sqrt{1-\mu_{1}^{2}}/\Delta^{1/4}$.
The transformation of variables needed to bring \MTsoln\ into the form
described by Eq.~(2.5) of \cite{LLM} includes \eqn{xyDef}{
\sqrt{x_1^2 + x_2^2} &= \mu_1\sqrt{r^2+L^2+q_1} \cr
y &= rL\sqrt{1-\mu_1^2} \,,
} and the all-important function $z$ of \cite{LLM} is \eqn{zDef}{
z = {1 \over 2} {r^2 + (q_1-L^2)(1-\mu_1^2) \over
r^2 + (q_1+L^2)(1-\mu_1^2)} \,.
} It is possible to use \xyDef\ to eliminate $\mu_{1}$ and $r$ and express $z$
solely in terms of $y$ and $\sqrt{x_{1}^{2}+x_{2}^{2}}$. The resulting
expression is somewhat lengthy, and we will not record it here. Its
$y\rightarrow0$ limit is quite simple, though: \eqn{zLimit}{
z\Big|_{y=0} = \left\{
\seqalign{\span\TC &\quad \span\TT}{
{1 \over 2} { q_1 - L^2\over q_1 + L^2}
& if $x_1^2 + x_2^2 < q_1 + L^2$  \cr
1/2 & otherwise.
} \right.
} If $q_{1}=0$, then we recover the free fermion description of $AdS_{5}\times
S^{5}$, which is $N_{c}$ fermions filling a disk-shaped droplet of radius $L$
in the $x_{1}$-$x_{2}$ plane. The one-charge superstar is some excitation of
this state, and from \zLimit\ we see that it is a uniformly but partially
filled disk of radius $\sqrt{q_{1}+L^{2}}$. The filling fraction must be
$\nu=L^{2}/(q_{1}+L^{2})$ in order for the number of fermions still to be
$N_{c}$. Comparing with \zLimit, we arrive at the result \eqn{NuToZ}{
z\Big|_{y=0} = {1 \over 2} - \nu \,.
} This partially filled state is not unique: it should be regarded as an
ensemble where the quantum states of individual fermions corresponding to
points inside the disk each have probability $\nu$ of being occupied. Any
representative of this ensemble may be constructed from the ground state by
acting with a color-singlet polynomial in the complex scalar field
$Z_{1}=X_{1}+iX_{2}$---the particular polynomial being constructed from
factors corresponding to Young diagrams in a manner that is described quite
precisely in \cite{Berenstein,LLM} following earlier work, notably \cite{RamgoGGs}. A rough approximation of this
construction is to think of many powers of $\det Z$ and sub-determinants of it
multiplied together to give the operator used for constructing a
representative state of the ensemble. These products are the CFT translation
of the distribution of giant gravitons found in \cite{SuperStars}.

It is interesting to write the thermodynamics of the near-extremal
single-charge superstar in terms of $\nu$: we recall from (\ref{entropysingle}%
) that \eqn{ESagain}{
S = {4\pi \over 3} (\Delta-J_1)
{\sqrt{2J_1/N_c^2} \over 1+2J_1/N_c^2} =
{4\pi \over 3} (\Delta-J_1) \sqrt{\nu(1-\nu)} \,,
} in a regime where $1\ll\Delta-J_{1}\ll J_{1}$. This is quite an interesting
regime because as one increases the \textquotedblleft twist\textquotedblright%
\ $\Delta-J_{1}$ from $0$, the singular BPS solution gradually grows a regular
horizon. Counting the number of states with large $J_{1}$ and moderate
$\Delta-J_{1}$ seems closer to tractable than other entropy problems
encountered in the study of strongly coupled $\mathcal{N}=4$ super Yang-Mills.
Perhaps surprisingly, the expression for the entropy in \ESagain\ involves no
explicit factors of $N_{c}$, encouraging us to think that a correct counting
of color singlets will provide a complete enumeration of the relevant microstates.

\section{Conclusions and outlook}

\label{CONCLUDE}

We have shown that the salient thermodynamic features of R-charged black holes
in $AdS_{5}$ can be understood in terms of effective strings created from the
intersection of giant gravitons. \ In the single charge case the free fermion
picture of \cite{LLM} make it possible to give a very explicit description of
the dual CFT operators, and we have briefly indicated in
section~\ref{FREEFERMIONS} how they can be constructed. \ In this concluding
section we consider the two and three charge cases. \ Letting $Z_{i}$ denote
the three complex scalars charged under $U(1)^{3}\subset SU(4)_{R}$, the
operators $\det Z_{i}$ each correspond to a maximal giant graviton with
angular momentum along a distinct equator of the $S^{5}$ \cite{DETDET}.

Upon identifying the two distributions of giant gravitons with operators built
from $Z_{1}$ and $Z_{2}$ fields, the effective string moves along the third
equator of the $S^{5}$ and thus consists of $N_{L}$ left-moving $Z_{3}$ fields
and $N_{R}$ right-moving $\bar{Z}_{3}$ fields in the dual CFT. The total
dimension of a corresponding gauge theory operator is then:%
\begin{equation}
\Delta=N_{L}+N_{R}+J_{1}+J_{2}+\delta_{imp}%
\end{equation}
where the presence of the additional small impurity term $\delta_{imp}$
follows from the fact that the operators must remain above the
BPS\ bound\footnote{See equation \ref{beeboba}.}. As of the writing of this
paper, we do not yet know of a systematic way to insert the $Z_{3}$ and
$\bar{Z}_{3}$ fields and the order $\delta_{imp}$ of other impurities and will
thus simply write the microstates of such black holes, $\left\{
\Omega_{\text{BH}}\right\}  $ as:%
\begin{equation}
\left\{  \Omega_{\text{BH}}\right\}  \sim\left\{  \overset{J_{1}/N_{c}%
}{\underset{i=1}{\prod}}\int\det X(x_{i})\rho\left(  x_{i}\right)  d^{4}%
x_{i}\overset{J_{2}/N_{c}}{\underset{j=1}{\prod}}\int\det Y(y_{j})\rho\left(
y_{j}\right)  d^{4}y_{j}\right\}  _{\substack{\text{all possible insertions of
}\\N_{L}\text{ }Z^{\prime}s\text{, }N_{R}\text{ }\bar{Z}^{\prime}%
s\\\delta_{imp}\text{ other impurities}}}
\end{equation}
where the presence of the $d=4$ distribution $\rho$ simply indicates that we
do not yet understand how localized the operators are on the CFT and\ the
product over determinants runs from $1$ to $J_{1}/N_{c}$ rather than
$2J_{1}/N_{c}$ because the determinant prescription we are using is still too
crude to correctly model angular distributions of giant gravitons on the
$S^{5}$. It would be interesting to explicitly verify that the entropy
generated by all such gauge theory operators correctly matches onto the known
black hole entropy. Finally, we recall that in contrast to the single charge
case where a black hole can form for an arbitrarily small value of $\mu>0$, in
both the double and triple charge black holes, the extremality parameter must
remain above the finite value $\mu_{\text{crit}}>0$. This behavior suggests
the possibility of a phase transition in passing from the single to higher
charge cases. This should correspond to a sufficiently large value of
$\delta_{imp}$ in the dual CFT, but we do not yet understand the dynamics
responsible for this phase transition. Perhaps it is analogous to a free
energy of crystallization, where the microstates of the black hole are like
the high-entropy, high-energy fluid phase and the BPS state is like the
crystalline phase.

As the above analysis demonstrates, even the smallest semiclassical R-charged
black holes must have R-charge at least $N_{c}$ in order for the intersecting
giant graviton picture to produce a physically reasonable effective string.
Although the effective strings for black holes in flat space and $AdS_{5}$ are
both produced from the intersection of D3-branes, there are some important
differences between the properties of these strings. Indeed, whereas black
holes in flat space are described by effective strings produced by order $J$
D3-branes, black holes in $AdS_{5}$ are described by effective strings
produced by only order $J/N_{c}$ D3-branes. \ The ubiquitous factors of
$N_{c}$ in both the supergravity formulas and effective string relations
underscore the key role that the five-form flux plays in the dynamics of
extended objects in $AdS_{5}\times S^{5}$. In any case, a more detailed
description of the dual CFT\ operators for such configurations would be most instructive.

\section*{Acknowledgements}

We thank O.~Lunin, J.~Maldacena, and J.~McGreevy for useful discussions. The
work of SSG~was supported in part by the Department of Energy under Grant
No.\ DE-FG02-91ER40671, and by the Sloan Foundation.

\appendix{}

\section*{Appendix}

In this appendix we compute the central charge of the effective string in the
two and three-charge cases by calculating the intersection number of two giant
graviton distributions sourced by charges $q_{1}$ and $q_{2}$. \ The case of
one giant graviton distribution intersecting a single transverse D3-brane
follows as a special case of the following analysis and will be omitted.
\ Using notation established in section \ref{EFFECTIVE}, we introduce the
directional cosines:%
\begin{equation}
\mu_{1}=\cos\theta_{1},\text{ }\mu_{2}=\sin\theta_{1}\cos\theta_{2},\text{
}\mu_{3}=\sin\theta_{1}\sin\theta_{2}.
\end{equation}
The self-dual RR five-form flux is then given by:%
\begin{equation}
F^{(5)}=dB^{(4)}+\ast dB^{(4)}%
\end{equation}
where \cite{SuperStars}:%
\begin{equation}
B^{(4)}=-\frac{r^{4}}{L}\Delta dt\wedge d^{3}\Omega-L\underset{i=1}%
{\overset{3}{\sum}}q_{i}\mu_{i}^{2}\left(  Ld\phi_{i}-dt\right)  \wedge
d^{3}\Omega
\end{equation}
with $d^{3}\Omega$ the volume element of a unit $S^{3}$ with angles
$(\alpha_{1},\alpha_{2},\alpha_{3})$. \ Since $\Delta$ only depends on the
radial coordinate $r$, we will not need its explicit form in the computation
to follow. \ Following \cite{SuperStars}, the first distribution of giant
gravitons sourced by charge $q_{1}$ generates a flux $F^{(5,1)}$ with
component along the $S^{5}$:%
\begin{equation}
F_{\theta_{1}\phi_{1}\alpha_{1}\alpha_{2}\alpha_{3}}^{(5,1)}=L^{2}q_{1}%
\sin2\theta_{1}\sin^{2}\alpha_{1}\sin\alpha_{2}.
\end{equation}
Integrating over all coordinates except $\theta_{1}$ yields the number
distribution of gravitons in the $\theta_{1}$ direction:%
\begin{equation}
dn_{1}=\frac{N_{c}}{4\pi^{3}L^{4}}d\theta_{1}\int F_{\theta_{1}\phi_{1}%
\alpha_{1}\alpha_{2}\alpha_{3}}^{(5,1)}d\phi_{1}d^{3}\alpha=\frac{2J_{1}%
}{N_{c}}\sin2\theta_{1}d\theta_{1} \label{numone}%
\end{equation}
where in the last equality we have used the dictionary entries previously
established. \ Similarly, the second distribution of giant gravitons sourced
by charge $q_{2}$ generates two components along the $S^{5}$:%
\begin{align}
F_{\theta_{1}\phi_{2}\alpha_{1}\alpha_{2}\alpha_{3}}^{(5,2)}  &  =-L^{2}%
q_{1}\sin2\theta_{1}\cos^{2}\theta_{2}\sin^{2}\alpha_{1}\sin\alpha_{2}\\
F_{\theta_{2}\phi_{2}\alpha_{1}\alpha_{2}\alpha_{3}}^{(5,2)}  &  =L^{2}%
q_{1}\sin^{2}\theta_{1}\sin2\theta_{2}\sin^{2}\alpha_{1}\sin\alpha_{2}.
\end{align}
The number distribution of giant gravitons due to the $q_{2}$ charge is then:%
\begin{equation}
dn_{2}=\frac{2J_{2}}{N_{c}}\left(  \sin^{2}\theta_{1}\sin2\theta_{2}%
d\theta_{2}-\sin2\theta_{1}\cos^{2}\theta_{2}d\theta_{1}\right)  .
\label{numtwo}%
\end{equation}
To determine the intersection number of these two distributions, we consider
the intersection of the infinitesimal distributions at angles $\theta_{1}$ and
$\theta_{2}$ so that the spheres intersect over some $S^{1}$. \ Now perform
the Hopf fibration along this $S^{1}$:%
\begin{equation}
\varphi:S^{5}\rightarrow S^{5}/S^{1}\simeq\mathbb{CP}^{2}.
\end{equation}
Since an $S^{3}$ inside of the $S^{5}$ is given by a degree 2 real algebraic
variety and since the giant graviton distribution at this angle is this same
$S^{3}$ wrapped $dn$ times, the first distribution of giant gravitons maps to
a complex curve $C_{1}(\theta_{1})$ in $\mathbb{CP}^{2}$ of degree $2dn_{1}$
and the second distribution of giant gravitons maps to a complex curve
$C_{2}\left(  \theta_{2}\right)  $ of degree $2dn_{2}$. \ The total
intersection number $I(\theta_{1},\theta_{2})$ of the curves $C_{1}(\theta
_{1})$ and $C_{2}(\theta_{2})$ is then given by B\'{e}zout's theorem, which
states that $I=\deg C_{1}\deg C_{2}$, or:%
\begin{equation}
I(\theta_{1},\theta_{2})=4dn_{1}dn_{2}\text{.}%
\end{equation}
To determine the total intersection number of the two distributions, we recall
that we are simply counting the number of units of flux from the five-form
flux generated by the giant gravitons. \ For this reason, we must integrate
the number distributions of equations (\ref{numone}) and (\ref{numtwo}) over
their positive domains of definition, $\theta_{1},\theta_{2}\in\lbrack
0,\pi/2]$. \ The total intersection number is thus:%
\begin{align}
I_{\text{tot}}  &  =\int_{0}^{\pi/2}\int_{0}^{\pi/2}I(\theta_{1},\theta
_{2})d\theta_{1}d\theta_{2}\\
&  =16\frac{J_{1}}{N_{c}}\frac{J_{2}}{N_{c}}\int_{0}^{\pi/2}\sin^{2}\theta
_{1}\sin2\theta_{1}d\theta_{1}\int_{0}^{\pi/2}\sin2\theta_{2}d\theta
_{2}=8\frac{J_{1}}{N_{c}}\frac{J_{2}}{N_{c}}\text{.}%
\end{align}
This $I_{\text{tot}}$ corresponds to the bosonic central charge of the string.
\ Assuming that the effective string has some supersymmetry, each worldsheet
boson has a fermionic counterpart so that the total central charge for the
effective string is:%
\begin{equation}
c_{\text{eff}}=\frac{3}{2}I_{\text{tot}}=12\frac{J_{1}}{N_{c}}\frac{J_{2}%
}{N_{c}}.
\end{equation}
\bibliographystyle{ssg}
\bibliography{adsbh}

\end{document}